\documentclass[pra,letterpaper,twocolumn]{revtex4-1}
\usepackage[utf8x]{inputenc}
\usepackage{amsmath}
\usepackage[T1]{fontenc}
\usepackage{graphicx}
\usepackage{dcolumn}
\usepackage{bm}
\usepackage{subfigure}
\usepackage{color}%
\usepackage{setspace}
\usepackage{amsfonts}%
\usepackage{amssymb}

\newcommand{\ket}[1]{\left| #1 \right\rangle}
\newcommand{\bra}[1]{\left\langle #1 \right|}

\newcommand{\abs}[1]{\left| #1 \right|}

\newcommand{\expn}[1]{{\rm e}^{#1}}

\newcommand{\eg}{\emph{e.g.,}~}
\newcommand{\ie}{\emph{i.e.,}~}
\newcommand{\nn}{\nonumber}

\makeatletter

\renewenvironment{widetext@grid}{%
  \par\ignorespaces
  \setbox\widetext@top\vbox{%
   \vskip15\p@
   \hb@xt@\hsize{%
    \leaders\hrule\hfil
    \vrule\@height6\p@
   }%
   \vskip6\p@
  }%
  \setbox\widetext@bot\hb@xt@\hsize{%
    \vrule\@depth6\p@
    \leaders\hrule\hfil
  }%
  \onecolumngrid
  \let\set@footnotewidth\set@footnotewidth@ii
}{%
  \par
  \twocolumngrid\global\@ignoretrue
  \@endpetrue
}%

\makeatother
\newcommand{\smallfrac}[2]{\mbox{ $\frac{#1}{#2} $ }}

\newcommand{\expect}[1]{\ensuremath{\left\langle{#1}\right\rangle}}

\newcommand{\dg}{^{\dagger}}

\newcommand{\half}{\smallfrac{1}{2}}

\makeatletter

\renewcommand\@make@capt@title[2]{%

\@ifx@empty\float@link{\@firstofone}{\expandafter\href\expandafter{\float@link}}%

{\textbf{#1}}\@caption@fignum@sep#2\quad}%

\makeatother

\makeatletter 

\renewcommand{\fnum@figure}{\textbf{Figure~\thefigure}}

\makeatother

\begin{document}


\onecolumngrid

\begin{flushleft}
\begin{spacing}{1.5}
{\textbf {\Large Fundamental Limits to Single-Photon Detection Determined by Quantum Coherence and Backaction}}
\end{spacing}

\vspace{\baselineskip}

{\textbf {\large Steve M. Young, Mohan Sarovar, Fran\c{c}ois L\'{e}onard}}

\vspace{\baselineskip}

Sandia National Laboratories, Livermore, CA, 94551, USA

\vspace{\baselineskip}

\end{flushleft}

\begingroup
\leftskip2em
\rightskip2em

{\noindent
\textbf{Single-photon detectors 
have
achieved impressive performance, and have led to a number of new scientific
discoveries and technological applications.  
Existing models of photodetectors are semiclassical in that the field-matter interaction is treated perturbatively and time-separated from
physical processes in the absorbing matter.    
An open question is
whether a fully quantum detector, whereby the optical field, the optical
absorption, and the amplification are considered as one quantum system,
could have improved performance. Here we  develop a theoretical model of such photodetectors and employ simulations  to reveal the critical role played by quantum coherence and amplification backaction in dictating the performance. We show that coherence and backaction lead to tradeoffs
between detector metrics, and also determine optimal system designs through
control of the quantum-classical interface. Importantly, we establish the design parameters that result in a perfect photodetector with 100\% efficiency, no dark counts,
and minimal jitter, thus paving the route for next generation detectors.}
}

\endgroup

\vspace{\baselineskip}

\twocolumngrid

Modern models of photodetectors and the photodetection process are rooted in pioneering work in quantum optics and quantum electronics \cite{Glauber:1962tt, Kelley:1964fn, Scully:1969gg, Ueda:1990iv,Mandel:1995ub}, and have not been significantly
modified or updated since. This is surprising given the degree to
which experimental photodetection technology has progressed over the past
century. Indeed, single-photon photodetectors have been developed based on a wide
range of physical processes that span from the photoelectric effect in
semiconductors \cite{Bienfang:2004ij,Woodson:2016cx} to superconductivity \cite{Pernice:2012bc,Marsili:2012ib,Marsili:2013th}, and moreover, these photodetectors have achieved impressive performance in terms of efficiency, dark
count rate, and jitter \cite{Eisaman:2011cc, Hadfield:2009}. Furthermore, advances in materials science and
nanoscale engineering open up possibilities for not only tuning the
microscopic properties and dynamics of photodetectors, but also to develop
entirely new classes of photodetectors. Such possibilities motivate a
re-examination of photodetection theory, with a view on
understanding the fundamental limits and tradeoffs. 
In particular, an open question is whether a photodetector where the electromagnetic field, light-matter interaction and amplification processes are all captured within a single quantum mechanical system could reveal new regimes of photodetector performance.

In this work we present such a re-examination of photodetection by developing a fully quantum mechanical minimal model for photodetection and examining the
fundamental limits that emerge from this model. Our approach relies on recent advances in quantum optics theory and quantum measurement theory. As shown in Fig.\ 1, the 
photodetection process consists of three components: (i)
coherent interaction between the electromagnetic (EM) field and a localized
system (usually some matter degrees of freedom), (ii) localization of
information about portions of the EM field state, usually through a transfer
of energy from the EM field to the localized system, and (iii) amplification
of this information to classical/macroscopic degrees of freedom. Importantly, 
we do not assume that the dynamics 
of each of these components are
necessarily at different timescales and thus effectively noninteracting. Such an
assumption is implicit in the traditional theory of photodetection and most
subsequent treatments that treat the light-matter interaction perturbatively. 

\begin{figure*}[ptb]
\setlength\belowcaptionskip{-10pt}
{\includegraphics[scale=0.8]{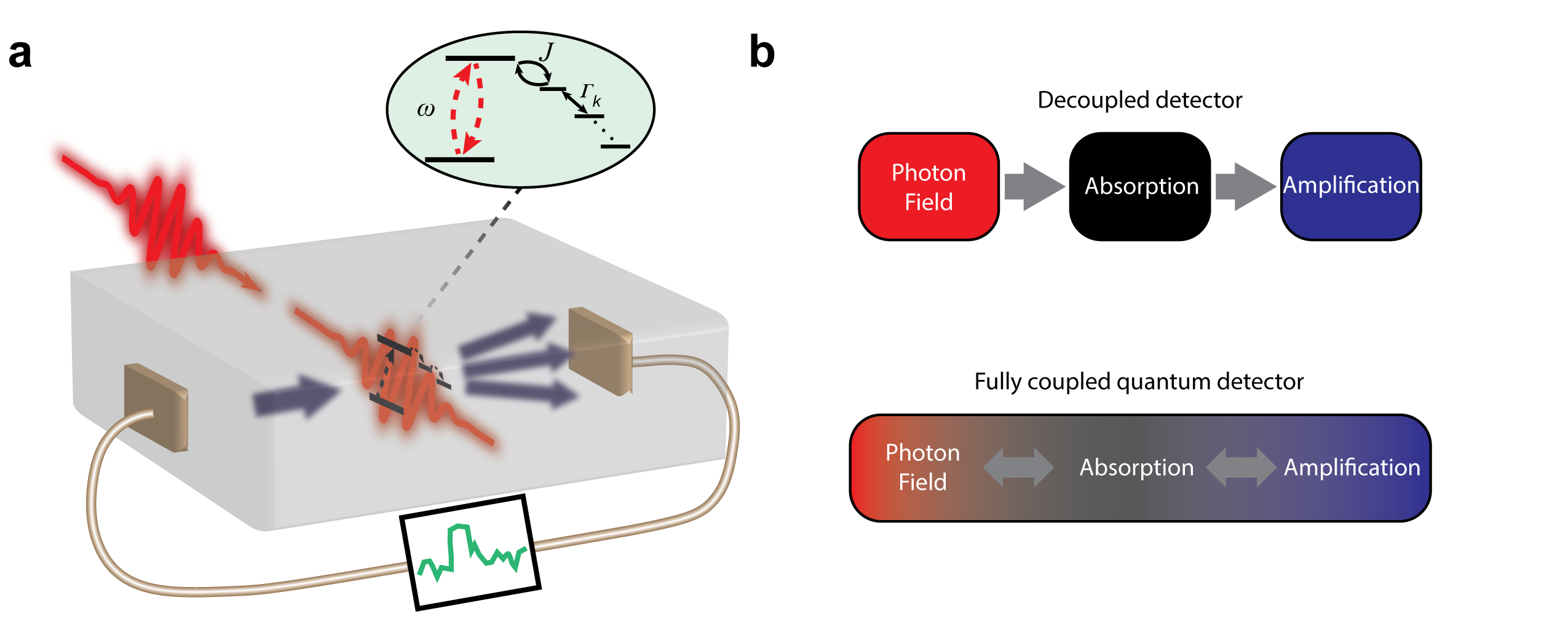}}
\caption{\textbf{Single-photon detection in a fully-coupled detector.} \textbf{a}, Illustration of a photodetector where photon wavepackets interact with the detector (matter) degrees of freedom. The optically coupled excited state can decay into a number of optically inactive states. Information carriers, \eg electrons, interact with the matter energy levels and scatter, causing
a response that can be monitored through a measurement. \textbf{b}, The conventional theory of photodetection assumes that the photon field, the absorption process,
 and the amplification process occur at different timescales and can therefore be treated separately. Alternatively, in this work we consider a fully coupled model where the three subsystems are treated as being part of one quantum system. }%
\label{fig:gen_model}%
\end{figure*}

In order to simulate the dynamics of photodetection we use an open quantum systems formalism \cite{Bre.Pet-2002}, and develop a master equation that explicitly accounts for the EM field degrees of freedom and internal degrees of freedom of the detector (the energy states depicted in Fig.\ \ref{fig:gen_model}). We use the
formalism developed by Baragioloa \emph{et al.} \cite{Baragiola:2012cs} for
modeling the interaction of few-photon wavepackets with matter.   This formalism is fully quantum mechanical
and therefore captures the modification of the field mode as a
result of  interaction with the detector.  The state of the detector internal degrees of freedom at any time is given by the density matrix $\varrho(t) = \sum_{MN}c^{*}_{MN}\rho^{MN}(t)$, where $c_{MN}$ are defined by an expansion of the initial state of the field in terms of Fock state wavepackets, \emph{i.e.} $\rho_{\rm field}(0) = \sum_{MN} c_{MN}\ket{M_{E}}\bra{N_{E}}$, with $\ket{N_E}$ being an $N$-photon wavepacket with temporal envelope given by $E(t)$, see Methods. $\rho^{MN}$ are auxillary density matrices whose temporal evolution is dictated by a hierarchy of coupled differential equations \cite{Baragiola:2012cs}. In our context, to model the three components of photodetection mentioned above, these coupled equations take the form: 
\begin{flalign}
\dot{\rho}^{MN}&(t)=\mathcal{M}(\rho^{MN}(t))+\mathcal{A}(\rho^{MN}(t))+\left\{\mathcal{D}[L]\left(\rho^{MN}(t)\right)\right.\nonumber\\
&+\sqrt{M}E(t)\left[\rho^{M-1N}(t),L^\dagger\right]\nonumber\\
&\left.+\sqrt{N}E^*(t)\left[L,\rho^{MN-1}(t)\right]\right\}\label{eq:general}
\end{flalign}
where $\mathcal{M}$ represents the dynamics of the detector internal degrees of freedom (that lead to excitation localization), $\mathcal{A}$ represents the dynamics of the amplification process, and the remaining terms capture the field-matter excitation dynamics ($L$ is the matter operator coupled to the field -- \eg a transition operator that couples the optically active states). The superoperator $\mathcal{D}$ is defined as $\mathcal{D}[A]\rho\equiv A\rho A^{\dagger}-\frac{1}{2}A^{\dagger}A\rho-\frac{1}{2}\rho A^{\dagger}A$. In the following we discuss the details of the superoperators $\mathcal{M}$ and $\mathcal{A}$; further details of this dynamical equation and its origins are presented in Methods. 

The optically active internal states of the detector are coupled to
a variable number of other states (that do not interact with the
EM field) either coherently or incoherently. These internal states could
represent \emph{e.g.,} excitonic or electronic states of a solid-state
material, or even electronic or conformational states of molecules. This coupling is
captured by $\mathcal{M}$, which describes the dynamics of the internal states that effectively localizes a photoexcitation within the detector degrees of freedom and funnels it away from the optically active state:
\begin{flalign}
\mathcal{M}\rho=-i[H,\rho]+\sum_{k}\Gamma_{k}\mathcal{D}[\left\vert e_{i_{k}%
}\right\rangle \left\langle e_{j_{k}}\right\vert ]\rho.
\end{flalign}
Here, $H$ is the Hamiltonian describing the energies of all internal states in
the device (denoted $\left\vert e_{l}\right\rangle $) and coherent couplings
between them; $\Gamma_{k}$ is the incoherent transition rate from state
$e_{j_{k}}$ to $e_{i_{k}}$. 
Any incoherent transitions are a result of interactions with reservoirs, \eg phonon degrees of freedom; we do not explicitly model these here, and instead capture their net effect on the essential internal states of the detector.

Finally, $\mathcal{A}$ represents the amplification component captured using quantum measurement theory.
A designated final internal state, $\left|  X \right\rangle $, is
continuously monitored, a process modeled using a quantum measurement master equation that can be derived from general principles  \cite{Cav.Mil-1987, Cresser:2006dc,Jac.Ste-2006}. 
This monitoring effectively amplifies information about occupation of that state by generating a classical measurement record that depends on the population of the state. 

The average effects of the amplification are captured by the term:
\begin{flalign}
\mathcal{A}\rho=\chi \mathcal{D}[\ket{X}\bra{X}]\rho
\label{eq:gen_mme}
\end{flalign}

with $\chi$ a rate
that quantifies how strongly the state is monitored, or equivalently the rate at which information about the internal states is being amplified into the classical domain. The associated average measurement current is given by
\begin{flalign}
\bar{I}_{t}=\int_{t-t_m}^{t}\chi^2\varrho_{XX}(t')dt',
\label{eq:avg_current}
\end{flalign}
where $t_m$ is the integration time window, and $\varrho_{XX}(t)$ is the population of the $X$ state given by the physical density matrix for the detector internal degrees of freedom, $\varrho(t)$. In our calculations, we chose $t_{m}$ values that result in optimal performance for a given $\chi$.
Such a Markovian description of the amplification 
process is not universal, but importantly, it 
captures the fact that any amplification process must have an associated backaction on the system being amplified \cite{Caves:1982zz,Clerk:2010dh}. An advantage to modeling the amplification process as a continuous measurement is that we can utilize quantum trajectory theory
\cite{Wiseman:2009vw,Jac.Ste-2006} to \textquotedblleft unravel'' a measurement master
equation into a stochastic master equation (SME) that
enables simulation of the photodetection system conditioned on particular
measurement records. This enables simulation of individual photodetection records and associated dynamics, in addition to the average record and dynamics given by Eqs. \ref{eq:gen_mme} and \ref{eq:avg_current}. The explicit form of the associated SME can be found in Methods. 

The above general model applies to a broad range of physical systems. In the Supplementary Information we examine an explicit
physical device for photodetection and show how one can derive a
description like the one used here from physical interaction
models.

\begin{figure}[ptb]
\setlength\belowcaptionskip{-10pt}
{\includegraphics[scale=0.4]{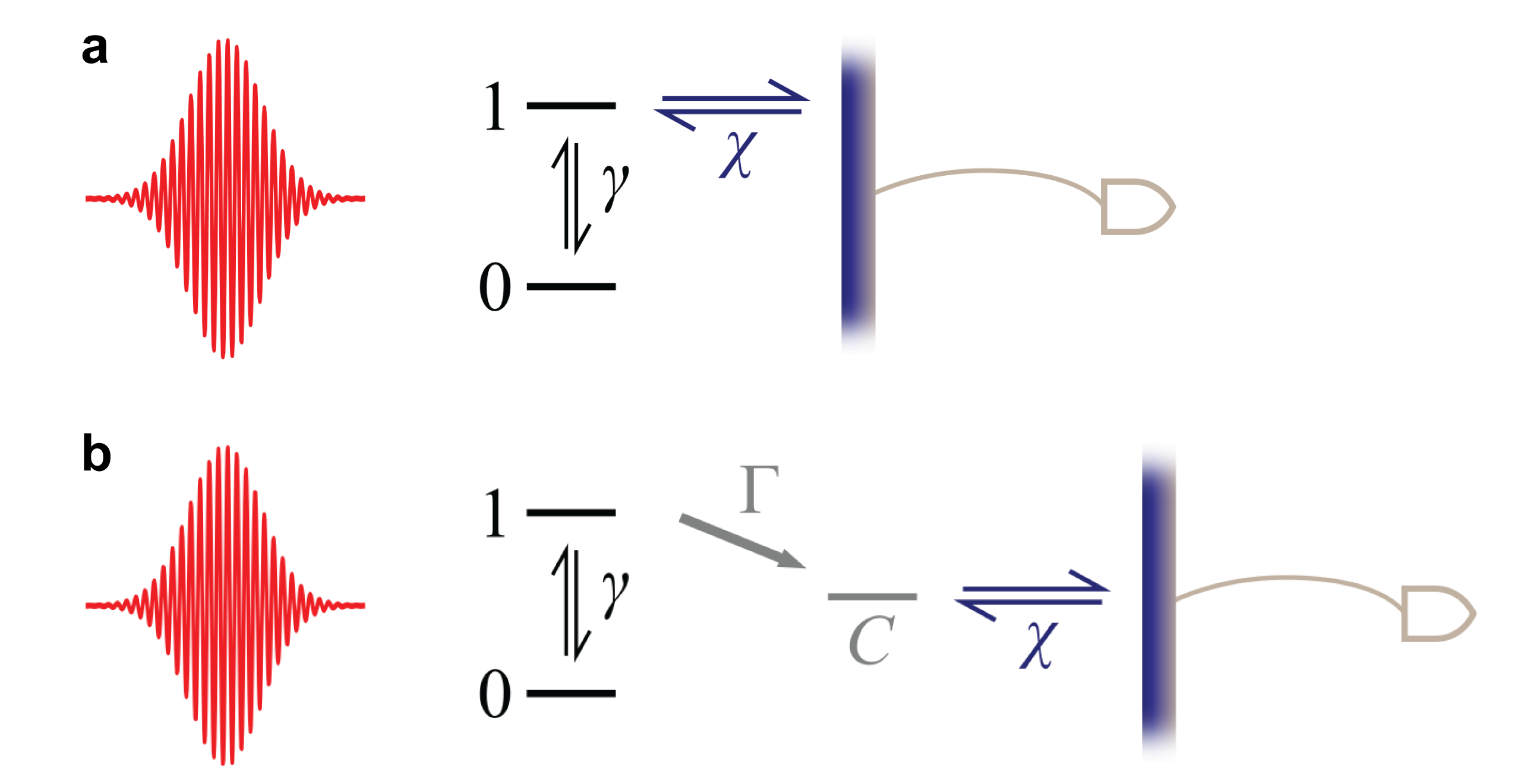}\label{fig:sys1}}
\caption{\textbf{The two configurations under consideration.} In both cases, we consider a
single-photon Gaussian wavepacket (left) incident on a two-level system, creating a
resonant excitation. A quantum measurement element (vertical purple line) couples to the internal states
and amplifies the signal to the classical domain (right). In \textbf{a}, the
quantum measurement element directly couples to the excited state and so $\ket{X} \rightarrow \ket{1}$, while in \textbf{b}, the population in the excited state may decay
into a third, optically inert state to which the quantum measurement element is coupled, and so $\ket{X} \rightarrow \ket{C}$.}%
\label{fig:sys}%
\end{figure}

Given the above minimal model of a photodetector and its dynamics, we ask several fundamental design questions: What is the best arrangement of internal states and couplings between them in order to maximize performance? Is a time-scale separation between the light-matter interaction and subsequent internal dynamics optimal? To answer such questions, it turns out to be sufficient to study two specific detector configurations: Configuration 1, with the final monitored state, $\ket{X}$, being the same as the optically active state (Fig.\ \ref{fig:sys}a), and Configuration 2, with $\ket{X}$ being a long-lived dark state, $\ket{C}$, to which the optically active state incoherently decays (Fig.\ \ref{fig:sys}(b)). 
In the latter configuration we assume that $\ket{1}$ and $\ket{C}$ are sufficiently separated in energy so that thermally excited population transfer from $\ket{C}$ to $\ket{1}$ can be neglected. Such a thermal effect can be modeled but would only yield a trivial decrease in efficiency of this configuration. 
In the following, we study both of these configurations using the dynamical model described above and quantify detector performance in terms of efficiency, dark count rates,  and jitter.

We assume that the incoming field contains a single photon with a Gaussian temporal profile 
$\abs{E(t)}^2=\left(\frac{1}{2\pi\sigma^2}\right)^{\frac{1}{2}}\expn{-\frac{t^2}{2\sigma^2}}$ of width $\sigma=1$ns. The field-matter coupling is $L=\gamma\ket{0}\bra{1}$ with $\gamma=0.01\mathrm{ps}^{-1}$, which produces near-maximal
absorption probability of $\approx 80\%$ for an isolated, unmonitored two-state system \cite{Wang:2011kn}. Since we consider an initial state containing a single photon, we need only propagate $\rho^{11}$, $\rho^{01}$, $\rho^{10}$ and $\rho^{00}$. 

\vspace{0.1in}

\textbf{\emph{Configuration 1.}}
The hierarchy of dynamical equations, Eq. \ref{eq:general}, for the system in Configuration 1 
can be written in component form as
\begin{flalign*}
	\dot{\rho}_{01}^{01}&=-i\omega_{01}\rho^{01}_{01}-\gamma E(t)-\frac{\gamma^2+\chi^2}{2}  \rho^{01}_{01}; \quad \rho_{01}^{01}(0) = 0, \\
	\dot{\rho}_{00}^{11}&=2\gamma E(t)\rho^{01}_{01}+\gamma^2\rho^{11}_{11}; \quad \rho_{00}^{11}(0)=1,\\
	\dot{\rho}_{11}^{11}&=-2\gamma E(t)\rho^{01}_{01}-\gamma^2\rho^{11}_{11}; \quad \rho_{11}^{11}(0)=0,
\end{flalign*} 
with $\rho^{00}_{00}=1$ throughout, and all other elements zero throughout.

The
generation of the coherence between states $\ket{0}$ and $\ket{1}$ is damped by both the
spontaneous emission back into the photon mode and decoherence due to
backaction from the amplification of $\ket{1}$; this restricts the development of the
excited state population. Solving the equations for different values of $\chi$
makes this concrete; stronger amplification noticeably reduces the
excitation probability (Fig.\ 3a),
a manifestation of the Zeno effect \cite{Mis.Sud-1977,Fac.Pas-2001}.

\begin{figure*}[hptb]
\setlength\belowcaptionskip{-10pt}
{\includegraphics[scale=0.73]{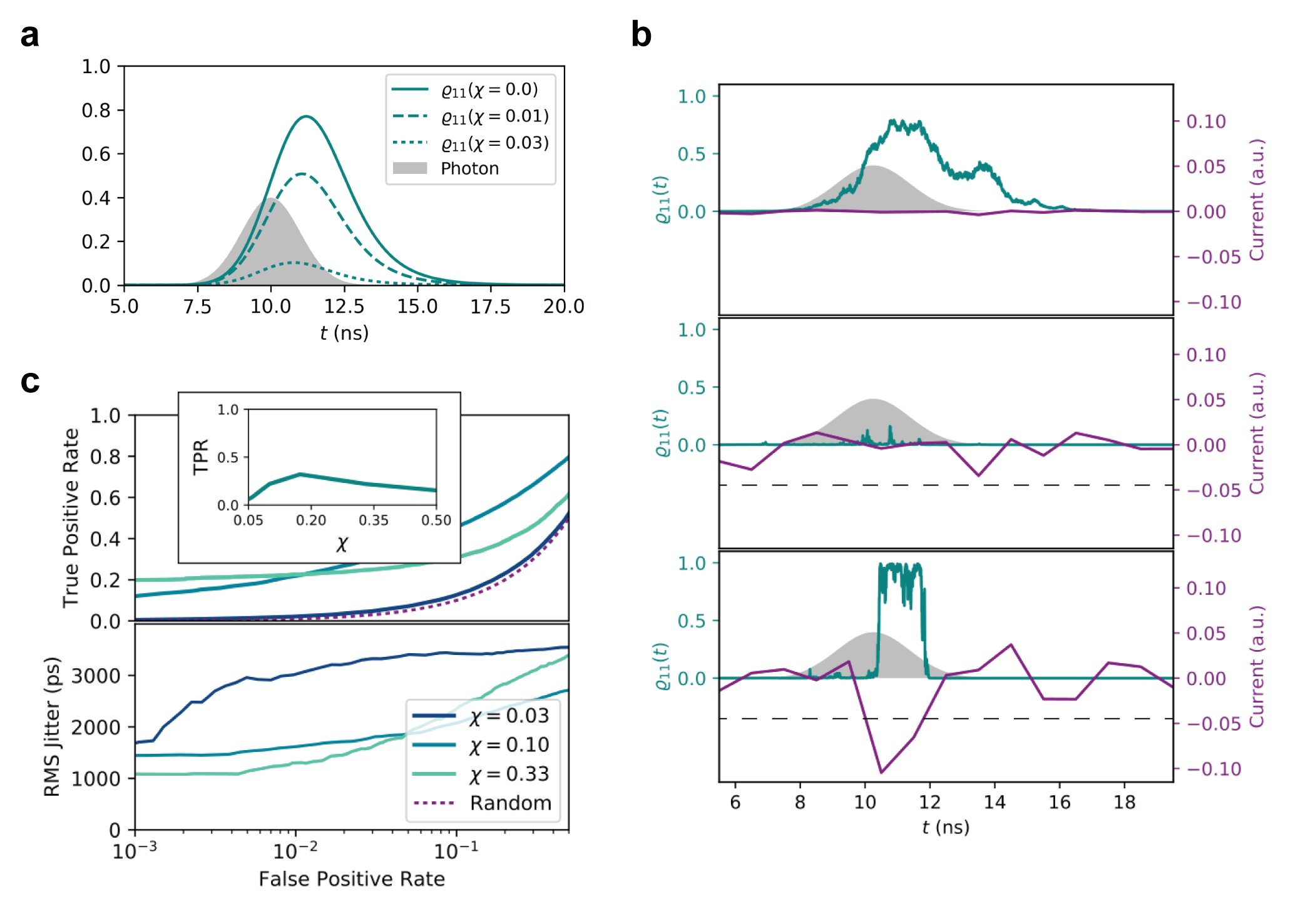}}
\caption{\textbf{Detection events and performance for Configuration 1.} \textbf{a}, The averaged population dynamics for Configuration 1 with amplification strengths
	 $\chi=0$, $\chi=0.01\mathrm{ps}^{-1}$,
	and $\chi=0.03\mathrm{ps}^{-1}$ .  The instantaneous current is proportional to $\chi^{2}\rho_{11}$. \textbf{b},
Sample trajectories and current outputs for $\chi=0.00333\mathrm{ps}^{-1}$ (top panel) and $\chi=0.0333\mathrm{ps}^{-1}$ (bottom two panels). The horizontal dashed lines are examples of current thresholds used to calculate the ROC curves. \textbf{c},  ROC curve and jitter vs. false positive rate,
for several amplification strengths obtained by computing each for varying
current thresholds for detection. The dashed line is the result for ``detectors'' that simply record random hits at varying rates, giving equal true positive and false positive rates. In the inset, the efficiency
for a false positive rate of 0.01 is plotted as a function of $\chi$; optimal
detection efficiency is obtained for intermediate amplification strength for a
modest rate of false positives and strong amplification strength for minimal false
positives. Stronger coupling also reduces jitter, which increases with the
false positive rate. }%
\label{fig:1D2L}%
\end{figure*} 

Simulating individual trajectories reveals additional aspects of the tradeoff between information gain and disturbance.
Figure 3b shows the excitation population
and associated detector output for sample trajectories. 
For weak amplification,
Fig.\ 3b, the individual trajectories are similar 
to the averaged case. Unfortunately,
due to the weak coupling the current cannot be readily distinguished
from the background noise. 
Stronger amplification significantly alters population evolution and 
produces trajectories that 
either completely miss or completely absorb the
photon. In this case the current unambiguously distinguishes photon absorption events, but at the price of reduced efficiency due to significant perturbation of the absorption probability. 

A more complete picture of impact on photodetection performance emerges after compiling the results over many
trajectories. Figure 3c shows the Receiver Operating
Characteristics (ROC) curve obtained from simulating 1000 trajectories with a photon and
1000 trajectories without a photon, for each value of $\chi$.
The true positive rate (TPR) corresponds to the fraction of trajectories when the detector output exceeds
a pre-defined threshold in the presence of a photon in the field, while the false positive rate (FPR) is when the detector output exceeds the
threshold without a photon being incident on the detector. Each point on the ROC curve is for
a different value of the threshold, which decreases from left to right. For large thresholds,
both the TPR and the FPR are low, while for low threshold both the TPR and the FPR are high.
We find a clear tradeoff between TPR and FPR regardless of the threshold used.

We also show in the inset of Fig.\ 3c the efficiency as a function of the amplification strength $\chi$ for a fixed FPR of 0.01. As might be anticipated from the average dynamics, the
efficiency is maximized for intermediate amplification strength: too weak and the signal cannot be reliably
separated from the noise, too strong and excitation is suppressed. 
Calculation of the RMS jitter (Fig.\ 3c) reveals no tradeoff with dark counts:
a low FPR is associated with low jitter. Unfortunately this occurs when the TPR (efficiency)
is low. At the higher efficiency levels, where intermediate coupling maximizes efficiency, we find that the same intermediate coupling also gives the lowest jitter.
\vspace{-0.03in}

Ultimately, we find that directly amplifying the optical excitation interferes with the excitation itself, creating a tradeoff between increasing the signal-to-noise ratio and avoiding amplification-induced decoherence.

\begin{figure*}[hptb]
\setlength\belowcaptionskip{-10pt}
{\includegraphics[scale=0.73]{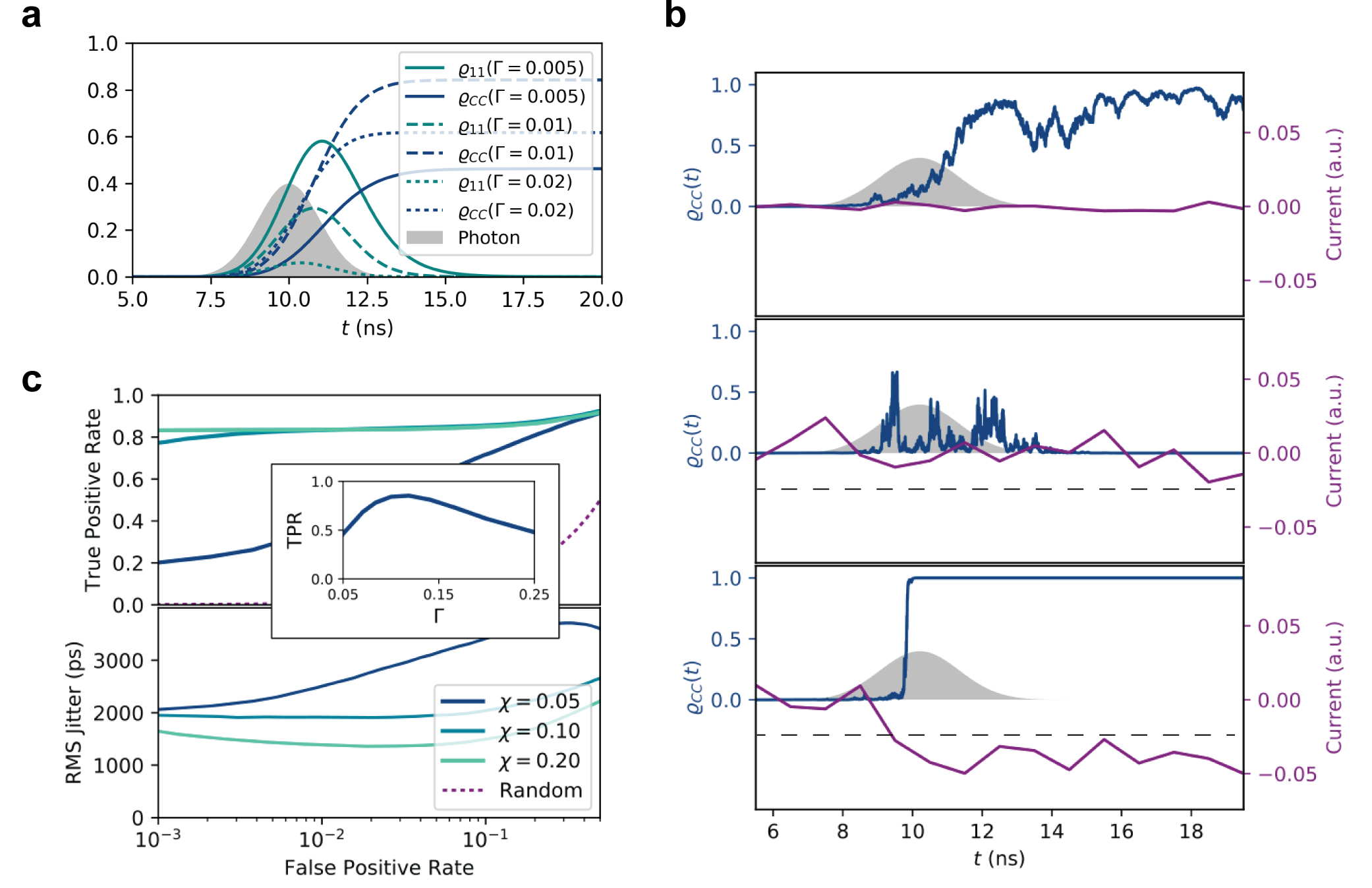}}
\caption{\textbf{Detection events and performance for Configuration 2.} \textbf{a}, Average population dynamics for diferent values of the incoherent transfer rate $\Gamma$.  \textbf{b}, Sample
trajectories and detector output for  $\chi=0.005\mathrm{ps}^{-1}$ (top panel) and for $\chi=0.02\mathrm{ps}^{-1}$ (bottom two panels) where the photon is
not absorbed and when the photon induces an
excitation. The horizontal dashed lines are examples of current thresholds used to calculate the ROC curves.
\textbf{c}, ROC curve and jitter vs. false positive rate
for several amplification strengths . Very high detection efficiency is obtained
for strong amplification at essentially no cost in terms of dark count
rate or jitter. The efficiency is plotted against the rate of relaxation into
the dark state $\Gamma$ (inset). There is a clear maximum; faster transfer collects
excitation more efficiently but inhibits the excitation process.
}%
\label{fig:1D3LR_st}%
\end{figure*}

\textbf{\emph{Configuration 2.}} In the case where the excited state decays to a dark state
(Fig.\ 2b), the matrix equations become 
\begin{flalign*}
	\dot{\rho}_{01}^{01}&=-i\omega_{01}\rho^{01}_{01}+\gamma E(t)-\frac{\gamma^2+\Gamma^2}{2}  \rho^{01}_{01}; \quad \rho_{01}^{01}(0)=0, \\
	\dot{\rho}_{00}^{11}&=2\gamma E(t)\rho^{01}_{01}+\gamma^2\rho^{11}_{11}; \quad \rho_{00}^{11}(0)=1,\\
	\dot{\rho}_{11}^{11}&=2\gamma E(t)\rho^{01}_{01}-(\gamma^2+\Gamma^2)\rho^{11}_{11}; \quad \rho_{11}^{11}(0)=0, \\
	\dot{\rho}_{CC}^{11}&=\Gamma^2\rho^{11}_{11}; \quad \rho_{CC}^{11}(0) = 0,
\end{flalign*} 
with $\rho^{00}_{00}=1$ throughout, and all other elements zero throughout. $\Gamma$ is the incoherent decay rate from $\ket{1}$ to $\ket{C}$. 
We note that the amplification strength appears nowhere in these equations. Thus,
in contrast to Configuration 1, the average dynamics exhibit no influence from
amplification strength and backaction. Instead, the coupling to the decay state introduces decoherence in a similar fashion as the amplification in Configuration 1. As such, it produces a similar tradeoff: there is an optimal value for the decay rate into this
state. As seen in Fig.\ 4a, a slow decay rate  allows for high
excitation probabilities but low population of the measured state, while fast
decay rates convert more of the excited
population into the measured-state population, but reduce the excitation
probability through decoherence. 

Despite the average dynamics being insensitive to the amplification strength, 
the relationship between information gain and disturbance is still affected by the details of the amplification. This is evident when examining individual trajectories (Fig.\ 4b).
The strong amplification 
yields currents that unambiguously signal absorption or non-absorption. Moreover, in this configuration, the long lifetime of the $\ket{C}$ state results in a persistent current when the photon has been absorbed.
Interestingly, although the amplification cannot influence the average populations (and hence there is no Zeno effect according to traditional definitions \cite{Fac.Pas-2001}), the amplification does effect the variance in the populations -- larger $\chi$ yields a larger variance in population statistics at a fixed time. 

Again, we can summarize the influence of various parameters by aggregate performance statistics
(Fig.\ 4c). The detection efficiencies are significantly
higher than for Configuration 1; moderate amplification
is sufficient to guarantee optimal efficiency with negligible dark counts and
no tradeoff must be negotiated. Indeed, Fig.\ 4c shows that for high threshold values, the TPR exceeds 0.8
while the FPR is 0.001. Similarly, jitter is much less sensitive to
the detection threshold. In contrast to Configuration 1, amplification does not adversely affect optical excitation; no tradeoff exists, and both efficiency and jitter are optimized by stronger,
rather than intermediate, amplification.  

In the above, we have taken the optical coupling to be in the
regime that provides optimal excitation probability for the isolated two state
system \cite{Wang:2011kn}. This optimum occurs due to the tradeoff between excitation rate and emission rate. However, the introduction of an
amplification mechanism adds both additional decoherence and protects against
emission back into the field mode. This suggests that the detector may be able
to take advantage of strong optical coupling. In addition, since our results indicate that
relaxation into the $\ket{C}$ state should occur at a similar rate as the excitation,
increasing the optical coupling means that the relaxation rate ought to be
increased as well, further enhancing decoherence and providing protection against
emission. 
Indeed, we find that, in
contrast to the un-monitored system, the optimal optical coupling is arbitrarily
high; the detector can actually achieve near perfect efficiency if both
$\gamma\gg1/\sigma$ and $\gamma=\Gamma$, as shown in Fig.\ 5. Essentially, the pulse is absorbed as quickly as possible and the resulting excited state population is shunted to the dark state as soon as it develops, preventing re-emission. 
Additionally, the amplification can be made arbitrarily strong, since the coherent field-matter interaction is decoupled from its backaction by the incoherent decay process, so
that dark counts can be essentially eliminated. Performing 50,000
simulations with and without a photon using $\gamma=\Gamma=0.1$ps$^{-1}$, we find that a wide range of
thresholds yield 50,000 hits, 0 dark counts, and 1.05ns of jitter, where 1.0ns
jitter is the lower limit set by the wavepacket width, $\sigma$.  Furthermore, almost perfect detection is possible
for a wide range of single-photon pulse widths
 -- \eg for the value $\gamma=\Gamma=0.1$ps$^{-1}$, such ideal performance holds for pulses as short as 100ps (see Supplementary Information for additional trajectories). 
For shorter pulses, efficiency is reduced (Fig.\ 5), as absorption no longer occurs rapidly enough to collect the entire pulse. 

This 
demonstrates that tradeoffs in photodetection can be 
circumvented through detector design, and that a perfect detector is in principle possible.

\begin{figure}[ptb]
\setlength\belowcaptionskip{-10pt}
\includegraphics[width=.95\columnwidth]{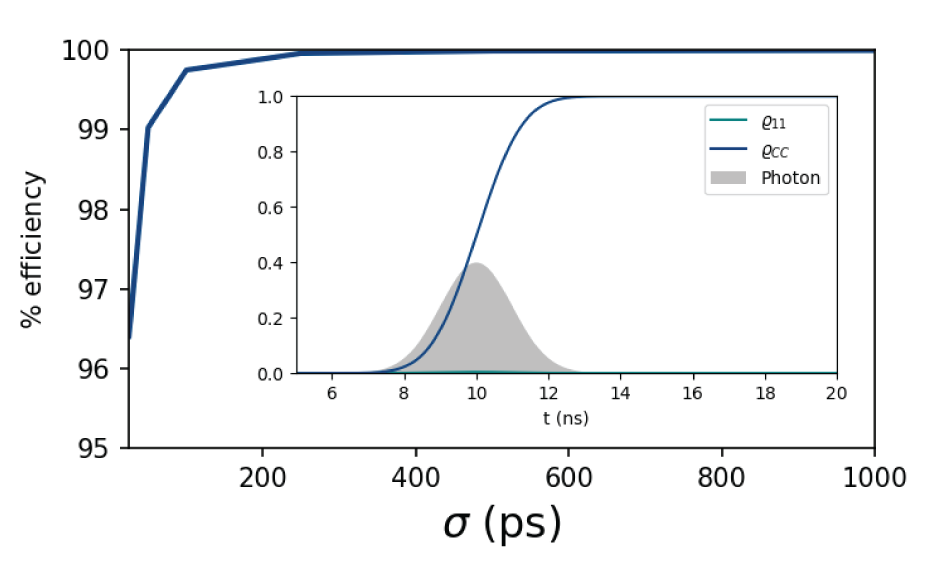}\label{fig:p100}
\caption{\textbf{Perfect photodetector.} The average dynamics of Configuration 2 with $\gamma
=\Gamma=0.1$ps$^{-1}$.  Strong optical coupling means that excitation occurs rapidly compared to the pulse duration, and a matched rate of incoherent transfer converts excited state population to dark state population as soon as the former develops.  Consequently, regardless of the strength of the quantum measurement, the measured 
state $\ket{C}$ attains 100\% probability and the excited state $\ket{1}$ is nearly unoccupied throughout.  }
\end{figure} 

At the most 
general level, the two models discussed here represent all the detectors of the kind presented in Fig.\ \ref{fig:gen_model}. 

  The presence of multiple decay states and/or extended decay pathways does not qualitatively alter our results, to the extent that final states are all monitored; decay processes to unmonitored states will obviously and straightforwardly limit performance.  
  
 Finally, we believe that 
current high-performance 
single-photon detectors should be described by Configuration 2.  
Both avalanche photodiodes and superconducting detectors rely on rapid incoherent decay from an optically excited coherent state to optically inert intermediate states before amplification.  Our results clearly show why these types of detectors offer the superior performance they are known for, and suggest that in principle this class of detectors may be tuned to operate perfectly, providing a fundamental design principle for guiding future efforts to engineer new photodetector types regardless of the underlying physical mechanisms.        

\vspace{\baselineskip}
\noindent{\large{\textbf{Methods}}}

To model the photon wavepacket and its interaction with the system, we adopt
the approach of Ref.~\cite{Baragiola:2012cs}. In the case of a propagating wavepacket with carrier frequency resonant with a localized system coupled to the field, 
the state of the localized system can be represented by the density matrix $\varrho(t) = \sum_{MN}c^{*}_{MN}\rho^{MN}(t)$, where $c_{MN}$ are defined by an expansion of the initial state of the field in terms of Fock state wavepackets, \emph{i.e.} $\rho_{\rm field}(0) = \sum_{MN} c_{MN}\ket{M_{E}}\bra{N_{E}}$, with $\ket{N_E} = 1/\sqrt{N!}\left[\int ds E(s) b^{\dagger}(s)\right]^N\ket{0}$ being an $N$-photon wavepacket (or continuous-mode Fock state) with temporal envelope given by $E(t)$. Under a Markov approximation for the field-matter interaction one may derive from the Ito Langevin equation a hierarchy of master equations for the auxiliary density matrices:

\begin{flalign*}
\dot{\rho}^{MN}(t)&=\mathcal{M}\rho^{MN}(t)+\mathcal{L}^{MN}_t\rho^{MN}(t)\\
&=\mathcal{M}+\mathcal{D}[L](\rho^{MN})+\sqrt{M}E(t)\left[\rho^{M-1N},L^\dagger\right]\\
&\hspace{3.5cm}+\sqrt{N}E^\dagger(t)\left[L,\rho^{MN-1}\right]
\end{flalign*}
\noindent
where $\mathcal{M}$ represents free evolution of the localized system and $L$ is an operator on the localized system that couples to the field. For all auxiliary density matrices where $N=M$, the initial state is the physical initial state of the localized system, while all other auxiliary density matrices are initialized to zero. 

The formalism developed in Ref.~\cite{Baragiola:2012cs} assumes the
field mode propagates along a 1D waveguide. The extension to 3D propagation 
may be done without much difficulty, but we do not consider it here. Also, the
resonant interaction assumption is not necessary, one could generalize to
interaction with several two-level systems of various detunings with respect
to the field mode.

The inclusion of amplification adds the term $\mathcal{A}$ to the master equations, so that in the case of
a single-photon wavepacket (where we only have auxiliary density matrices up to $N=M=1$) and real electric field $E(t)$, we have
\begin{flalign*}
\dot{\rho}^{11}&=-i[H,\rho^{11}]+ E(t)\left([\rho^{01},L^{\dagger}]+[L,\rho^{10}]\right)\\
&\hspace{2cm}+\mathcal{D}[L]\rho^{11}+\chi^2\mathcal{D}[\ket{X} \bra{X}]\rho^{01}\\
\dot{\rho}^{01}&=-i[H,\rho^{01}]+ E(t)[L^\dagger,\rho^{00}]\\
&\hspace{2cm}+\mathcal{D}[L]\rho^{01}+\chi^2\mathcal{D}[\ket{X} \bra{X}]\rho^{01}\\
\dot{\rho}^{00}&=-i[H,\rho^{00}]+\mathcal{D}[L]\rho^{00}+\chi^2\mathcal{D}[\ket{X} \bra{X}]\rho^{01}
\end{flalign*}
where $H$ is the free Hamiltonian for the system.

These equations describe the average or unmonitored state of the system. The
conditional dynamics, in which the detector output influences the
subsequent system dynamics, are obtained by the addition of the nonlinear term \cite{Wiseman:2009vw}
\begin{flalign*}
\chi\mathcal{H}&[\ket{X} \bra{X}]\rho^{MN}dW(t)& \\
&\equiv  \chi\Big(\ket{X} \bra{X}\rho^{MN} + \rho^{MN}\ket{X} \bra{X} \\ 
&~~~~~~~~~~~~~~ -2\langle \ket{X} \bra{X}\rangle_{\varrho(t)} \rho^{MN} \Big)\frac{dW(t)}{dt},
\label{eq:Hop}
\end{flalign*}
to each of the evolution equations in the hierarchy, resulting in a set
of stochastic differential equations. $dW(t)$ is a Wiener increment (Gaussian distributed random variable with mean zero and variance $dt$). We note that the
expectation value $\langle \ket{X}\bra{X} \rangle_{\varrho(t)}$ is evaluated under the physical density matrix $\varrho(t)$; this expectation value
is used to evaluate $\mathcal{H}[\ket{X}\bra{X} ]\rho^{MN}$ for the
evolution of the auxiliary density matrices as well. As a consequence, only
the trace of the real density matrix is conserved; the traces of the auxiliary
matrices are not. This is necessary to correctly reflect the conditioning of
the density matrix outcome on the measurement process.

The observed current consistent with this state evolution is given by
\begin{flalign*}
I_{t}=\int_{t-t_m}^{t}\chi^2\varrho_{XX}(t')dt' + \chi dW(t').
\end{flalign*}

We numerically solved the above equations using the order (2.0,1.5) stochastic Runge-Kutta algorithm proposed in Ref. \cite{Rossler:2010}. We perform simulations of the output signal when a photon is, or is not, incident
on the detector, and for a given threshold value of integrated current,
record hits, misses, dark counts, and jitter. For each set of parameters
considered we performed 1000 simulations with incident photons and 1000 without, from which we extract
true positive rates (percentage of the time a hit is recorded when a photon is
present), false positive rates (percentage of the time the current exceeds the
threshold when a photon is absent), and jitter (standard deviation of the time
a hit is recorded with respect to photon arrival time). In the case of the perfect detector
we extended these simulations to 50000 events in order to observe any extremely rare probability events.

\bibliographystyle{naturemag}
\bibliography{stoch_draft}

\vspace{\baselineskip}

{\setlength{\parindent}{0cm}
{\large
\textbf{Acknowledgments}
}

Work supported by the DARPA DETECT program. Sandia National Laboratories is a
multimission laboratory managed and operated by National Technology and
Engineering Solutions of Sandia, LLC., a wholly owned subsidiary of Honeywell
International, Inc., for the U.S. Department of Energy's National Nuclear
Security Administration under contract DE-NA-0003525.
}

\clearpage

\begin{widetext}
\begin{flushleft}
\begin{spacing}{1.5}
{\textbf {\Large Supplementary Information: Fundamental Limits to Single-Photon Detection Determined by Quantum Coherence and Backaction}}
\end{spacing}

\vspace{\baselineskip}

{\textbf {\large Steve M. Young, Mohan Sarovar, Fran\c{c}ois L\'{e}onard}}

\vspace{\baselineskip}

Sandia National Laboratories, Livermore, CA, 94551, USA

\vspace{\baselineskip}

\end{flushleft}

\noindent
{\textbf {\large 1. Example derivation of master equation from physical system}}

\vspace{\baselineskip}

In this section we derive a stochastic master equation representing a concrete physical system to illustrate the general formalism presented in the main text.
Figure \ref{fig:single_fermion} shows a schematic of the ``device" for which we will write a dynamical model. It consists of a photoactive molecule positioned near a short-channel ((\eg 10nm) carbon nanotube (CNT) connected to leads. There are three relevant states in the molecule: the ground state $\ket{0}$, an excited state $\ket{1}$ optically connected to the ground state, and an optically dark state $\ket{C}$ that represents the state of the molecule after photon absorption, such as can arise from photoisomerization. The static dipole of the molecule is different (in magnitude and direction) in states $\ket{1}$ and $\ket{C}$ and this change induces a change in the electrostatic potential of the CNT and as a result, the current across it. Such
systems have been previously studied experimentally \cite{Zhou2009}.

\begin{figure}[hbt]
  \includegraphics[scale=0.7]{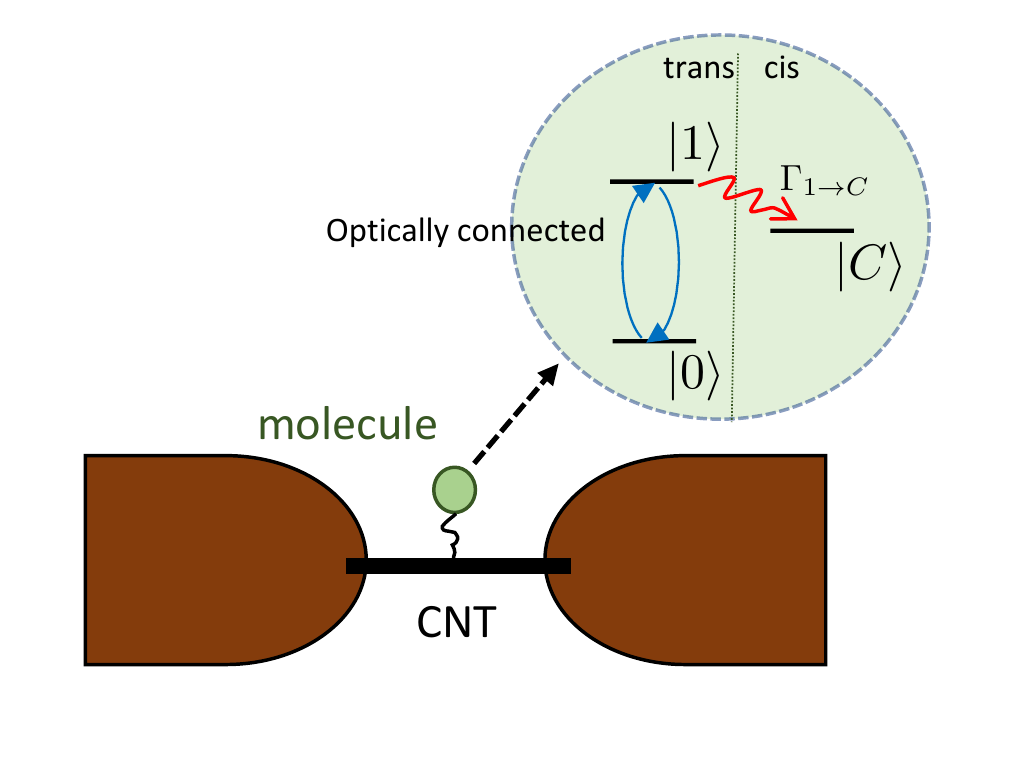}
  \caption{A schematic of a molecular photodetection device. Photons are absorbed by the molecular system (internal structure shown), and the detection event is amplified by electronic transport in the carbon nanotube. A quantum measurement master equation for the dynamics of this device is developed in this section.\label{fig:single_fermion}}
\end{figure}

In order to develop a model for this device, we will approximate the transport of electrons across the nanotube as a tunneling process, which is justified by the short channel length. The tunnel barrier is set by the energy levels in the nanotube. Given this simplification, we can model the system using a Hamiltonian  $H = H_f + H_{f-m} + H_m + H_{m-CNT} + H_{CNT}$, where $H_f, H_m$ and $H_{CNT}$ are the bare electromagnetic (EM) field, molecule and CNT Hamiltonians, respectively. $H_{f-m}$ is the field-molecular interaction, and $H_{m-CNT}$ is the molecule-CNT interaction. Explicitly, these Hamiltonians are:
\begin{flalign}
H_m &= \hbar\omega_0 \ket{0}\bra{0} + \hbar\omega_1\ket{1}\bra{1} + \hbar\omega_C \ket{C}\bra{C} \nn \\
H_{CNT} &= \sum_k (\hbar\omega_k^L a_{Lk}\dg a_{Lk} + \hbar\omega_k^R a_{Rk}\dg a_{Rk} + \sum_{k,q} (T_{kq}a_{Lk}\dg a_{Rq} + T^*_{qk}a_{Rq}\dg a_{Lk}) \nn \\
H_{m-CNT} &= \sum_{k,q} \ket{C}\bra{C}(\chi_{kq}a_{Lk}\dg a_{Rq} + \chi^*_{qk}a_{Rq}\dg a_{Lk}).
\end{flalign}
We will leave $H_f$ and $H_{f-m}$ unspecified at this point. In the above, $\hbar\omega_0,\hbar\omega_1,\hbar\omega_C$ are the energies of the respective states of the molecule. $\hbar\omega_k^L$ and $\hbar\omega_k^R$ are the energies of left and right reservoir/lead states at wavenumber $k$, and $a_{Lk}$, $a_{Rk}$ are (fermionic) annihilation operators for these states. $T_{kq}$ is the tunneling matrix element between states $k$ and $q$ in the left and right reservoir, and $\chi_{kq}$ is the perturbation to this element due to the molecule being in state $\ket{C}$. Note that this effectively means that the tunneling amplitude goes from $T_{kq}$ to $T_{kq}+\chi_{kq}$ when the molecule is in the $\ket{C}$ state. 

In addition to these coherent dynamics, the different conformational states $\ket{1}$ and $\ket{C}$ are connected by an incoherent rate $\Gamma_{1\rightarrow C}$ (and we assume the backward transition rate $\Gamma_{C \rightarrow 1}$ is negligible). 

This model for the CNT and molecule-CNT interaction are similar to the model used for quantum point contact based measurement in Ref. \cite{Goan:2001gb}. Following that reference, we can now derive a master equation describing the dynamics of the molecule and light degrees of freedom only by integrating out the continuum of reservoir states:
\begin{flalign}
\dot{\varrho}(t) &= -\smallfrac{i}{\hbar}[H_{f} + H_{f-m} + H_m,\varrho] + \Gamma_{1\rightarrow C}\mathcal{D}[\ket{C}\bra{1}]\varrho(t) + \mathcal{D}[\mathcal{T}_+ +\mathcal{X}_+ \ket{C}\bra{C}]\varrho(t)	+ \mathcal{D}[\mathcal{T}^*_- +\mathcal{X}^*_- \ket{C}\bra{C}]\varrho(t) \nn \\
&\equiv -\smallfrac{i}{\hbar}[H_{f} + H_{f-m} + H_m,\varrho] + \mathcal{L}_{t}\varrho
\label{eq:master_eq}
\end{flalign}
where $\varrho$ is the density matrix for the molecular and field degrees of freedom only. In this equation, $\mathcal{D}$ is a superoperator defined as:
\begin{flalign}
\mathcal{D}[A]\rho = A\rho A\dg - \half A\dg A \rho - \half \rho A\dg A	
\end{flalign}

Before specifying the coefficients $\mathcal{T}_{\pm}$ and $\mathcal{X}_{\pm}$ we repeat from Ref. \cite{Goan:2001gb} all the assumptions that go into deriving this master equation:
\begin{enumerate}
	\item The left and right reservoirs/leads are thermal equilibrium free electron baths.
	\item Weak coupling between molecule and CNT, which effectively means that we can restrict ourselves to a second order expansion in $\chi_{kq}, T_{kq}$.
	\item The transport through the channel (CNT/QPC) is in the tunnel junction limit -- \ie low transmmitivity.
	\item The initial state of the molecule and CNT are uncorrelated/factorizable.
	\item Fast relaxation of the reservoirs -- \ie the degrees of freedom in the reservoirs relax to equilibrium much faster than any system timescales.
	\item Markovian approximation of the reservoir.
	\item If $eV$ is the external bias applied across the transport channel, and $\mu_L, \mu_R$ are the chemical potentials in the left and right reservoirs (\ie $eV = \mu_L - \mu_R$), then $|eV|, k_BT \ll \mu_{L(R)}$.
	\item Energy independent tunneling amplitudes and density of states over the bandwidth $\textrm{max}(|eV|, k_bT)$.
\end{enumerate}
Under these approximations, the dynamics of the system is described by the above master equation, with the coefficient determined by
\begin{flalign}
	|\mathcal{T}_{\pm}|^2 &= \smallfrac{2\pi e}{\hbar} |T_{00}|^2 g_L g_R V_{\pm} \nn \\
	|\mathcal{T}_{\pm} + \mathcal{X}_{\pm}|^2 &= \smallfrac{2\pi e}{\hbar} |T_{00} + \chi_{00}|^2 g_L g_R V_{\pm},
\end{flalign}
where $T_{00},\chi_{00}$ are the energy-independent tunneling amplitudes near the average chemical potential, $g_L,g_R$ are the energy-independent density of states in the left and right reservoirs, respectively. The finite temperature effective external bias is:
\begin{flalign}
	eV_{\pm} \equiv \frac{\pm eV}{1-\exp\left(\frac{\mp eV}{k_BT}\right)}
\end{flalign}

At first approximation, we can work in the limit of low temperature and ignore the thermally activated current in the reverse direction, and set $V_-=0$, which will effectively remove the third term in \ref{eq:master_eq}.

Eq. \ref{eq:master_eq} can be interpreted as a measurement master equation giving the averaged dynamics when the population in the state $\ket{C}$ is continuously monitored \cite{Goan:2001gb}. Conditioned dynamics, based on particular values of the current can also be derived from the corresponding stochastic master equation \cite{Wiseman:2009vw}:
\begin{flalign}
d\varrho(t)	= -\smallfrac{i}{\hbar}[H_{f} + H_{f-m} + H_m,\varrho]dt + \mathcal{L}_{t}\varrho dt + \mathcal{H}[\mathcal{T}_+ +\mathcal{X}_+ \ket{C}\bra{C}]\varrho(t) dW_+(t) + \mathcal{H}[\mathcal{T}^*_- +\mathcal{X}^*_- \ket{C}\bra{C}]\varrho(t) dW_-(t), \nn
\end{flalign}
where $\mathcal{H}[A]\varrho \equiv A\varrho + \varrho A\dg - \expect{A+A\dg}_{\varrho}\varrho$, and $dW_+(t)$ and $dW_-(t)$ are Wiener increments. Increments in the forward and reverse current consistent with this evolution are given by:
\begin{flalign*}
	dI_+(t) &= \expect{\mathcal{T}_+ +\mathcal{X}_+ \ket{C}\bra{C}}_{\varrho}dt + dW_+(t), \nn \\
	dI_-(t) &= \expect{\mathcal{T}^*_- +\mathcal{X}^*_- \ket{C}\bra{C}}_{\varrho}dt + dW_-(t), \nn
\end{flalign*}

\vspace{\baselineskip}

\noindent
{\textbf {\large 2. Quantum trajectories for perfect detector}}
In this section we present additional trajectories for the "perfect photodetector" for different values of the photon wavepacket width. Fisure S2 shows that
for pulse widths ranging from 100ps to 1ns the collection efficiency is 100\%.

	\begin{figure*}[h]
		\includegraphics[scale=1.0]{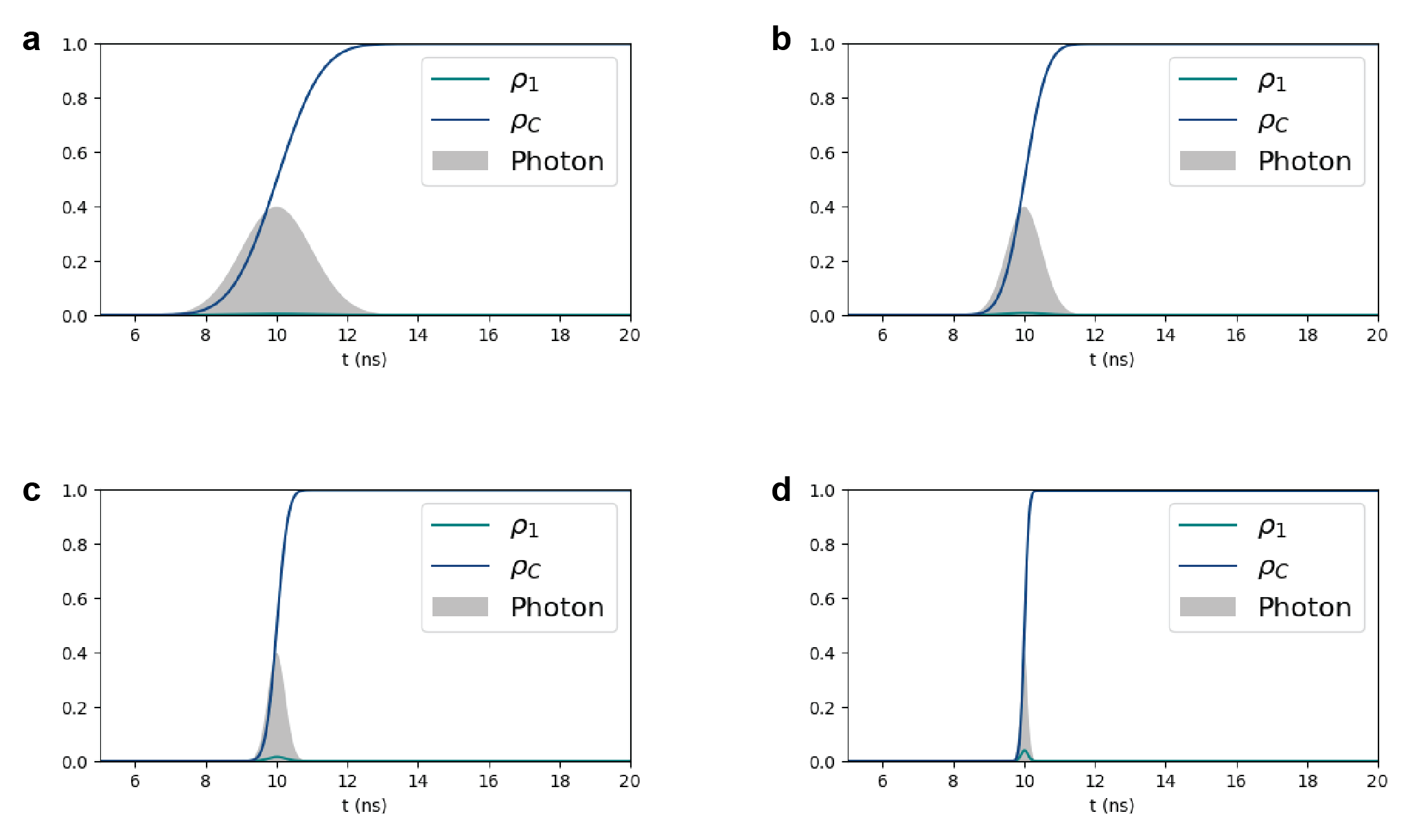}
		\caption{The average dynamics of the tree state system with $\gamma=\Gamma=0.1$ps$^-1$ for wavepackets of widths \textbf{a}, 1ns, \textbf{b}, 500ps, \textbf{c}, 250ps, and \textbf{d}, 100ps.  We see that in all cases the photon is collected 100\% of the time. } \label{fig:perfect}
	\end{figure*}

\end{widetext}	

\end{document}